\documentclass[a4paper,english,12pt]{article}
\usepackage[T1]{fontenc}
\usepackage[latin1]{inputenc}
\usepackage{babel}
\usepackage{graphics}

\makeatletter

\providecommand{\LyX}{L\kern-.1667em\lower.25em\hbox{Y}\kern-.125emX\@}
\newcommand{\noun}[1]{\textsc{#1}}

\makeatother
\begin{document}

\title{\textbf{Detection of Cosmic Shear from STIS Parallel Archive Data:
Data Analysis}}

\author{J.-M. Miralles\( ^{1,2} \), H. Hämmerle\( ^{2,3} \), N. Pirzkal\( ^{1} \),
P. Schneider\( ^{2,3} \),\and T. Erben\( ^{2,4,5} \), R.A.E Fosbury\( ^{1} \),
W. Freudling\( ^{1} \), B. Jain\( ^{6} \), S.D.M. White\( ^{3} \)}

\date{{\small \( ^{1} \)ST-ECF, Karl-Schwarzschild Str.2, D-85748 Garching
b. M\"unchen, Germany \\ \( ^{2} \)Institut f\"ur Astrophysik und
Extraterrestrische Forschung der Universit\"at Bonn, Auf dem H\"ugel 71,
D-53121 Bonn, Germany \\ \( ^{3} \)Max-Planck-Institut f\"ur Astrophysiks,
Karl-Schwarzschild Str.1, D-85741 Garching b. M\"unchen, Germany \\ \( ^{4}
\)Institut d'Astrophysique de Paris, 98bis Boulevard Arago, F-75014 Paris,
France \\ \( ^{5} \)Observatoire de Paris, DENIRM 61, Avenue de l'Observatoire,
F-75014 Paris, France \\ \( ^{6} \)Department of Physics and Astronomy,
University of Pennsylvania, 209 S. 33rd Street, Philadelphia, PA 19104
USA}}

\maketitle

\section{Introduction}

In June 1997, parallel observations using the Space Telescope Imaging
Spectrograph (STIS) on the HST started to be taken in substantial
numbers along many different lines-of-sight. We are using the imaging
data to investigate the distortion of background galaxies by the gravitational
field of the large scale matter distribution, also known as Cosmic
Shear. This effect was recently detected from the ground (Van Waerbeke
et al. 2000, Bacon et al. 2000, Kaiser et al. 2000, Maoli et al. 2001,
Wittman et al. 2000 and Van Waerbeke et al. 2001) and from space (Rhodes
et al. 2001). The typical object sizes that have to be measured are
in the order of < 0.5$''$. Therefore, STIS is perfectly suited
to such studies, thanks to its resolution and sensitivity. Also, due
to intrinsic cosmic variance, this project requires many observations
of separate fields, each containing tens of small faint background
galaxies, for which the parallel observations are adequate. This poster
presents the data and the catalog production that leads to the cosmic
shear result presented in poster \char`\"{}First Cosmic Shear results
from STIS parallel program archive data\char`\"{} (H\"ammerle et al. in this conference). The data is publicly
available also at http://www.stecf.org/projects/shear.

\section{\noun{D}ata \noun{}Reduction}

STIS CCD images provide a good deepth, excellent resolution and adequate
sampling of the telescope PSF. The detector has a pixel size of 0.05$''$
and a field of view of 51$''$ and is sensitive to wavelengths
from 2500 to 11000 Å. making it more efficient than WFPC2 for Cosmic
Shear studies.

The data used is a subset of the available STIS Parallel Survey Data
between June 1997 and October 1998 which satifies the following conditions: 

\begin{itemize}
\item Taken in the CLEAR filter mode (unfiltered CCD) 
\item CR-SPLIT mode 
\item Unbinned 
\item Associated \char`\"{}jitter ball\char`\"{} rms value smaller than
1/10 of a STIS pixel
\end{itemize}
Individual exposures were then associated and co-added using the procedure
described in Pirzkal et al. (2001). This procedure removes cosmic
rays, hot-pixels and uses a cross-correlation technique combined with
drizzling to achieve an accuracy of 1/10 of a pixel in the co-addition.
This procedure was tested through simulations of STIS data, confirming
that the shape and flux of individual objects was preserved.

498 co-added images were produced from which we identified 122 galaxy
fields (with more than 10 extended objects) and 55 star fields (with
more than 100 point like objects). The images on the right show you
a few exemples of galaxy and star fields. 

\begin{figure}
\resizebox*{0.85\textwidth}{!}{\includegraphics{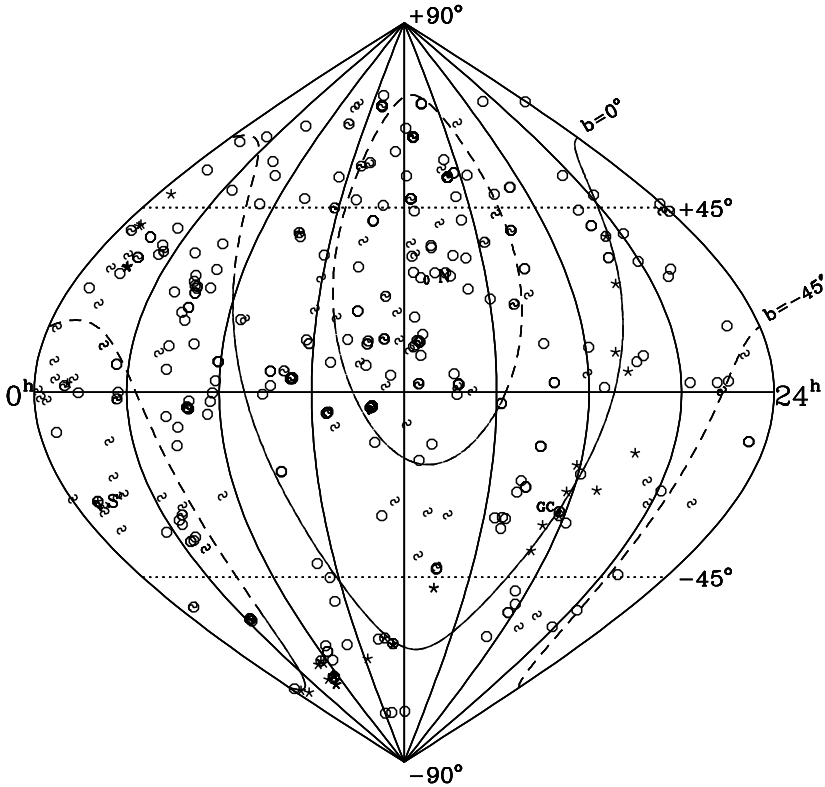}} 
\caption{Galactic coordinates positions of the 498 co-added fields. Stars, swirls and circles represent stellar fields, galaxy fields and non-classified fields.}
\end{figure}

\section{Data properties}

The properties of the 122 galaxy fields were studied more carefully
to characterize the objects observed: 

\begin{itemize}
\item The number counts of galaxies are consistent with previous estimates
(Gardner \& Satyapal 2000). 
\item The optimal integration time is between 2000 and 2500 seconds, leading
to an average number of 29 galaxies to be detected at 3 s level. 
\item The limiting magnitude reached, for a 5 pixel detection at 3 s level
is M$_{AB}$=28.5 in a 3600s exposure. 
\item The average size of galaxies with magnitudes ranging from 22 to 26
vary for 0.3$''$ to 0.1$''$ and are also consistent with
previous observations.
\end{itemize}
While the redshift distribution of these galaxies is currently unknown,
it is going to be determined using photometric redshift with VLT images
that are presently being obtained.

\vspace{0.3cm}
\begin{figure}
{\centering \resizebox*{0.7\textwidth}{!}{\includegraphics{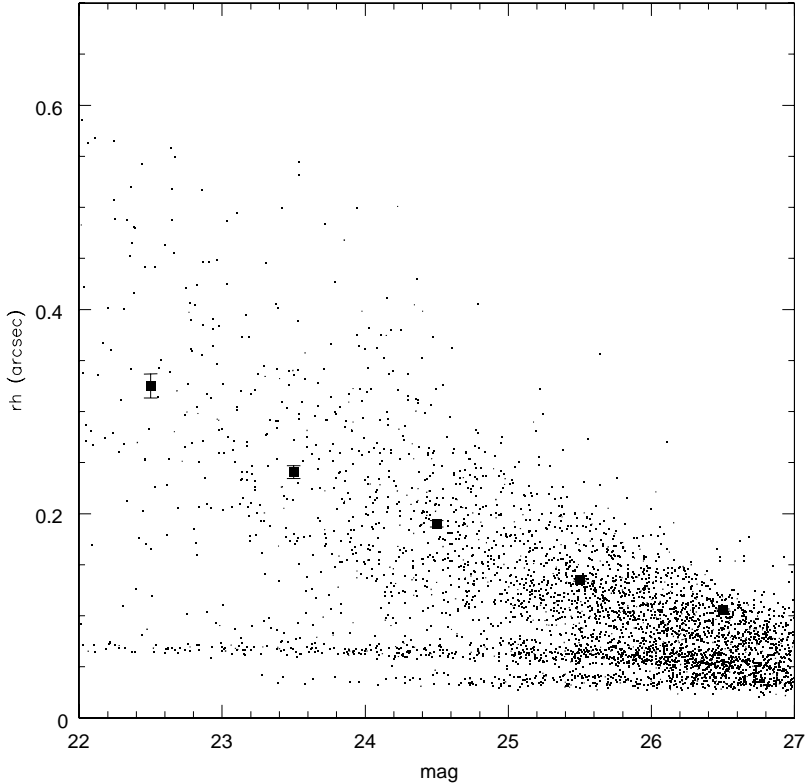}} \par}
\caption{Measured half-light radius (rh) of objects in co-added STIS images. The horizontal distribution of points at rh between 0.05" and 0.08" is caused by stars and other unresolved objects. The average sizes of galaxies per magnitude bin are indicated by square dot marks (counting only rh>0.08"). Errors bars represent the 1s level in the error of the mean.}
\end{figure}
\vspace{0.3cm}

\section{Catalog production}

Catalogs are produced using Sextractor (Bertin \& Arnouts 1996) and
IMCAT (Kaiser, Squires and Broadhurst 1995) following the subsequent
procedure: 

\begin{itemize}
\item Sextractor detects sources and calculates magnitudes 
\item IMCAT detects sources and measures sizes and shapes (e.g.: rh, e1
and e2) 
\item The final catalog is the result of the merging of the Sextractor and
IMCAT catalogs according to their coordinates. Only objects which
have an unique detection within a radius of 0.125$''$ in both
catalogs are kept (using the Sextractor coordinates as reference).
\end{itemize}

\section{For the Future...}

More STIS parallel data is actually being collected through a cycle
9 GO parallel program especifically dedicated to Cosmic Shear (Prop.
8562+9248, P.I.: P. Schneider). Since the end of September 2000 and up to date,
about 400 associations have been already produced and will allow us
to increase the accuracy and significance of the Cosmic Shear signal detected atscales less than the arcminute.

\vspace{0.5cm}
\textbf{Acknowledgments}: We wish to thank Ludovic Van Waerbeke, Yannick
Mellier, Emmanuel Bertin, Doug Clowe and Lindsay King for long, cheerful and
fruitful discussions. We also thank Alberto Micol and Stella Seitz
for their contributions to the project.

\end{document}